\documentstyle[aaspp4,12pt]{article}

\lefthead{J.J.M. in 't Zand et al.}
\righthead{GRB~980329 and its X-ray afterglow}

\begin{document}

\slugcomment{Accepted for publication in ApJ Letters on July 28, 1998}
\title{GAMMA-RAY BURST 980329 AND ITS X-RAY AFTERGLOW}
\author{J.J.M.~in~'t~Zand\altaffilmark{1,2}, 
 L.~Amati\altaffilmark{3,4},
 L.A.~Antonelli\altaffilmark{5,6},
 R.C.~Butler\altaffilmark{7},
 A.J. Castro-Tirado\altaffilmark{8,9},
 A.~Coletta\altaffilmark{10},
 E.~Costa\altaffilmark{3},
 M.~Feroci\altaffilmark{3},
 F.~Frontera\altaffilmark{11,12},
 J.~Heise\altaffilmark{1},
 S.~Molendi\altaffilmark{13},
 L.~Nicastro\altaffilmark{14},
 A.~Owens\altaffilmark{15},
 E.~Palazzi\altaffilmark{11},
 E.~Pian\altaffilmark{11},
 L.~Piro\altaffilmark{3},
 G.~Pizzichini\altaffilmark{11},
 M.J.S.~Smith\altaffilmark{1,10},
 M.~Tavani\altaffilmark{13,16}}
\affil{}
\altaffiltext{1}{Space Research Organization Netherlands, 
	Sorbonnelaan 2, 3584 CA Utrecht, the Netherlands}
\altaffiltext{2}{email jeanz@sron.nl}
\altaffiltext{3}{Istituto di Astrofisica Spaziale (CNR), 00133 Rome, Italy}
\altaffiltext{4}{Istituto Astronomico, Universit\`{a} Degli Studio "La Sapienza", Via Lancisi 29, 00100 Roma, Italy}
\altaffiltext{5}{Osservatorio Astronomico di Roma, Via Frascati 33, 00040 Monteporzio Catone, Italy}
\altaffiltext{6}{BeppoSAX Science Data Center, Via Corcolle 19, 00131 Rome, Italy}
\altaffiltext{7}{Agenzia Spaziale Italiana, Viale Regina Margherita, 00162 Roma, Italy}
\altaffiltext{8}{Laboratorio de Astrof\'{\i}sica Espacial y F\'{\i}sica Fundamental
      (INTA), P.O. Box 50727, 28080 Madrid, Spain}
\altaffiltext{9}{Instituto de Astrof\'{\i}sica de Andaluc\'{\i}a (CSIC),
      P.O. Box 18080, 18080 Granada, Spain}
\altaffiltext{10}{BeppoSAX Scientific Operation Center, Via Corcolle 19, 00131 Rome, Italy}
\altaffiltext{11}{Istituto di Tecnologie e Studio delle Radiazioni Extraterrestri (CNR), Via Gobetti 101, 40129 Bologna, Italy}
\altaffiltext{12}{Dipartimento Fisica, Universit\'{a} di Ferrara, Via Paradiso 12, 44100 Ferrara, Italy}
\altaffiltext{13}{Istituto Fisica Cosmica e Tecnologie Relative (CNR), Via Bassini 15, 20133 Milan, Italy}
\altaffiltext{14}{Istituto Fisica Cosmica e Applicazioni all'Informatica (CNR), Via Ugo La Malfa 153, 90146 Palermo, Italy}
\altaffiltext{15}{Astrophysics Division, Space Science Department of ESA, ESTEC, P.O. Box 299, 2200 AG Noordwijk, the Netherlands}
\altaffiltext{16}{Columbia Astrophysics Laboratory, Columbia University, New York,
                  NY 10027, U.S.A.}

\begin{abstract}
GRB~980329 is the brightest gamma-ray burst detected so far with the Wide Field 
Cameras aboard BeppoSAX, both in gamma-rays and X-rays. With respect to its 
fluence ($2.6\times10^{-5}$~erg~s$^{-1}$cm$^{-2}$ in 50 to 300 keV) it would be
in the top 4\% of gamma-ray bursts in the 4B catalog (Meegan et al. 1998).
The time-averaged burst spectrum from 2 to 20 and 70 to 650 keV can be well 
described by the empirical model of Band et al. (1993). The resulting 
photon index 
above the break energy is exceptionally hard at $-1.32\pm0.03$. An X-ray afterglow
was detected with the narrow-field instruments aboard BeppoSAX 7~h after 
the event within the error box as determined with the Wide Field Cameras. Its
peak flux is $(1.4\pm0.2)\times10^{-12}$~erg~s$^{-1}$cm$^{-2}$ (2 to 10 keV). The 
afterglow decayed according to a power law function with an index of 
$-1.35\pm0.03$. GRB~980329 is characterized by being bright and hard, 
and lacking strong spectral
evolution.
\end{abstract}



\section{Introduction}
\label{secintro}

After its launch in April 1996, the Wide Field Camera (WFC) instrument on 
board the BeppoSAX satellite opened an important window to the understanding 
of the gamma-ray burst (GRB) phenomenon. Through quick (within a few hours)
and accurate (within a few arcminutes) localizations it has enabled for the 
first time quick (within less than a day) and sensitive multi-wavelength 
follow-up observations of a number of GRBs. This has resulted in first time
detections of afterglows in a broad wavelength range. Two optical counterparts
have revealed redshifts which, if interpreted as due to the expansion of
the universe, place them at cosmological distances (Metzger et al. 1997, 
Kulkarni et al. 1998). 

Up to July 22, 1998, WFC
provided arcminute positions for 14 GRBs. The 
All-Sky Monitor (ASM) on board the Rossi X-ray Timing Explorer 
(e.g., Levine et al. 1996) provided such quick 
positions in 3 additional cases. Of these 17 bursts,
13 were followed up within a day in X-rays with sensitivities down
to roughly 10$^{-13}$~erg~s$^{-1}$cm$^{-2}$ in 2 to 10 keV. All of these
resulted in good candidates for X-ray afterglows.
Most of the afterglows decayed in a manner consistent with power law functions with
indices ranging from $-1.1$ for GRB~970508 (Piro et al. 1998) to
$-1.57$ for GRB~970402 (Nicastro et al. 1998). In two cases 
(GRB~970508, Piro et al. 1998, and GRB~970828, 
Yoshida et al. 1998) extra variability on top of the power law decay
was observed.

We here discuss the tenth GRB localized with BeppoSAX-WFC, GRB~980329. 
The quick distribution of the position (Frontera et al. 1998a, 
In~'t~Zand et al. 1998) resulted in the detection of a variable radio
counterpart that peaked about 3 days after the burst (Taylor et al. 1998).
Observations of the position of this radio source revealed a candidate host 
galaxy (Djorgovski et al. 1998) and fading coincident counterparts in the I band
(Klose et al. 1998), K band (Larkin et al. 1998, Metzger et al. 1998) 
and R band (Palazzi et al. 1998). The R-band counterpart decayed in a manner
consistent with a power law decay with an index of
$-1.3$. The optical magnitudes suggest a red spectrum which Palazzi
et al. (1998) suggest may be due to substantial absorption in a
starburst galaxy.

GRB~980329 is an exceptional case within the set of WFC-detected bursts: 
it is the brightest at gamma-rays as well as X-rays. Thus, GRB~980329 
provides an opportunity to probe previously unexplored parts of the
parameter space. We here present gamma-ray and X-ray 
measurements, both of the burst event and the afterglow of GRB~980329.

\section{Observations}
\label{sectionobs}

All GRB and X-ray afterglow measurements presented here were obtained with 
three sets of instruments on board BeppoSAX (Boella et al. 1997a).
The Gamma-Ray Burst Monitor (GRBM, Feroci et al. 1997, Costa et al. 1998) 
consists of the 4 lateral shields of the Phoswich
Detector System (PDS, Frontera et al. 1997)
and has a bandpass of 40 to 700 keV. The normal 
directions of two shields are each co-aligned with the viewing direction of a 
WFC unit. Therefore, a WFC-detected GRB has a near to optimum
GRBM collecting area. The GRBM
has 4 basic data products per shield for a GRB: a time history of the 40 to 
700 keV intensity with a variable time resolution of up to 0.48~ms, 
1~s time histories in 40 to 700 and $>$100 keV, and a 256 channel
spectrum accumulated each 128~s (independently phased from GRB
trigger times; 240 of these channels contain scientific data up to 650~keV).

The WFC instrument (Jager et al. 1997) consists of two coded aperture cameras 
each with a field of view of 40$^{\rm o}$ by 40$^{\rm o}$ full-width to zero 
response and an angular resolution of about 5\arcmin. The bandpass is 2 to 26 
keV. 

The narrow field instruments (NFI) include 2 imaging instruments that combined 
are sensitive to 0.1-10 keV photons: the low-energy and the medium-energy 
concentrator spectrometer (LECS and MECS respectively, see Parmar et al.
1997 and Boella et al. 1997b respectively). They both have
circular fields of view with diameters of 37\arcmin\ and 56\arcmin. 
The other 2, non-imaging, NFI are the PDS (13 to 300 keV),
and the high-pressure gas scintillation proportional counter (4 to 120
keV, Manzo et al. 1997).

GRB~980329 triggered the GRBM
on March 29.15587, 1998 UT. The peak intensity is $6.6\times10^3$~c~s$^{-1}$ in
40 to 700 keV. Simultaneously, the burst was detected in WFC unit number 2 at an 
off-axis angle of 19$^{\rm o}$. The peak intensity was about 6 Crab units in 2 to 
26 keV. This is the brightest burst seen with both GRBM and WFC. Nevertheless, the 
statistical quality of the WFC data is moderate because of the relatively large 
off-axis angle. The X-ray counterpart to GRB~980329 was localized with the WFC 
with an error circle radius of 3\arcmin\ (Frontera et al. 1998a). The burst also 
triggered BATSE on board CGRO (trigger number 6665, Briggs et al. 1998) and was 
detected above 1 MeV with the COMPTEL instrument on CGRO (Connors et al. 1998).

GRB~980329 was declared a target of opportunity for the 
NFI and a follow-up observation was started on March 29.4499 UT, 
7~h after the burst. The afterglow, 1SAX J0702.6+3850, was quickly 
identified in the LECS and MECS (In~'t~Zand et al. 1998). The 
time span of the observation is 41.5~h. An interruption occurred between 
21.2 and 27.5~h after the start. The total net exposure times are
25 ks for the LECS and 64 ks for the MECS. 

\section{Analysis}
\label{sectionana}

\subsection{The burst event}
\label{subsectionburst}

The most intense part of the burst has similar durations in X-rays and gamma 
rays (see figure~\ref{figlc}). The full-width 
at half-maximum duration of the burst is, within 0.5~s, equal to 8.6~s from 2
to 700 keV. The average burst duration versus photon energy relation as 
defined by Fenimore et al. (1995) for bright bursts predicts a time scale 
at 10 keV that is 2$\frac{1}{2}$ times larger that at 100 keV. 
There is a soft tail in the WFC data not easily discernible in figure~\ref{figlc}
but showing up in binned imaging data up until 90~s after the trigger (see 
figure~\ref{figth}). Beyond 40 keV the 
burst shows a moderate spectral evolution with a softening trend. 
The photon index as derived from a power law fit to the two-channel 1~s GRBM data 
(figure~\ref{figlc}c) varies between $-1.19\pm0.06$ and $-1.96\pm0.26$.
A Fourier power spectrum of the high time resolution data of the GRBM shows 
no features between $3$~Hz and 64~Hz.

The 256-channel GRBM data of GRB~980329 permits a sensitive spectral analysis.
A 128~s accumulation interval ends 54~s after the burst trigger and covers
most of the burst (see figure~\ref{figlc}). We have
analyzed this spectrum above 70 keV simultaneously with the 2 to 20 keV spectrum 
measured with WFC from 0 to 54~s (below 70 keV the response matrix of the GRBM is 
not sufficiently well known, the same applies to the WFC matrix above 20 keV). 
We employed the GRB spectral model defined by
Band et al. (1993) as:
\[ N(E) = \left\{ \begin{array}{ll}
                  AE^\alpha {\rm exp}(-E/E_0), & \mbox{for $E \leq (\alpha-\beta)E_0$,} \\
	          BE^\beta,                    & \mbox{for $E \geq (\alpha-\beta)E_0$ }
	          \end{array}
	  \right. 
\]
phot~s$^{-1}$cm$^{-2}$keV$^{-1}$, where $E$ is the photon energy and $A$ and $B$ 
are coupled normalization 
constants. A fit of this model to the data (see figure~\ref{figgrbspectrum}) 
reveals $\alpha=-0.6\pm0.2$, $\beta=-1.32\pm0.03$ and $E_0=134\pm110$~keV
($\chi^2_{\rm r}=1.07$ for 140 d.o.f.).
A broken power law function fits the data just as well (with comparable 
power law indices). Since $\beta$ is larger
than -2, even in the time-resolved data (see figure~\ref{figlc}c), the peak of 
the $\nu F_\nu$ spectrum is beyond the high energy boundary of our data set.
A single power law model with absorption due to cold
matter of cosmic abundances describes the data worse, with $\chi^2_{\rm r}=1.35$ 
(141 d.o.f.). The chance probability that $\chi^2_{\rm r}$
is at least that high is $2\times10^{-3}$.
Although the WFC and GRBM instruments have not yet been cross calibrated,
we anticipate no large relative systematic errors between WFC and GRBM. 
WFC could be calibrated for this particular observation with a
standard calibration source (i.e., the Crab source) in the same field of view. 
Furthermore, GRBM was calibrated against the Crab source and cross calibrated 
against BATSE for a number of bursts.

Given the model spectrum, the fluence is $(7\pm1)\times10^{-7}$~erg~cm$^{-2}$ in 2 to 
10 keV and $(5.5\pm0.5)\times10^{-5}$~erg~cm$^{-2}$ 
in 40 to 700 keV. The former fluence constitutes 1\% of the expected
2--700 keV fluence. The 50--300 keV fluence of
$(2.6\pm0.3)\times10^{-5}$~erg~cm$^{-2}$ would place GRB~980329 in the top
4\% of the bursts in the 4B catalog (Meegan et al. 1998).

\subsection{The X-ray afterglow}

The average LECS/MECS spectrum of 1SAX J0702.6+3850 over the first 21~h of the 
follow-up observation (up until the interruption) could be modeled by a power 
law function with a photon index of $\Gamma=-2.4\pm0.4$ and a low energy cutoff 
equivalent to interstellar absorption by a hydrogen column density of 
$N_{\rm H}=(1.0\pm0.4)\times10^{22}$~cm$^{-2}$ ($\chi^2_{\rm r}=0.9$ for 28 d.o.f.). 
The 99\% confidence interval for $N_{\rm H}$ is $0.21-2.3\times10^{22}$~cm$^{-2}$. A 
black body spectrum also fits the data well, the parameters are: 
k$T=1.0\pm0.1$~keV, $N_{\rm H}<2.4\times10^{22}$~cm$^{-2}$ (99\% confidence level) 
and $\chi^2_{\rm r}=1.1$ for 28 d.o.f. An f-test does not rule against
either spectral model. The average flux is 
$4.0\times10^{-13}$~erg~s$^{-1}$cm$^{-2}$ in 2 to 10 keV. The predicted Galactic 
value for $N_{\rm H}$ in the direction of the burst is $9\times10^{20}$~cm$^{-2}$ 
(interpolated from the maps by Dickey \& Lockman 1990). This is inconsistent 
with the NFI result on $N_{\rm H}$ for the power law model. The value of 
$1.9\times10^{21}$~cm$^{-2}$ as implied by the reddening of the optical spectrum 
(Palazzi et al. 1998) is marginally consistent with this value (the chance
probability for two $N_{\rm H}$ values being at least as deviant as observed
is 0.3\%). The data
do not allow a sensitive timing analysis of the spectrum. The roughest such 
analysis reveals that the 3 to 10 keV over 2 to 3 keV hardness ratio changes 
from $1.0\pm0.2$ to $0.5\pm0.2$ over the two halves of the observation.
This is statistically consistent with a constant ratio.

The 13 to 300 keV PDS data of the first 21~h of the follow-up observation show 
no detection of 
the afterglow. Assuming the same model spectrum applies as to the LECS and MECS
data, the 3$\sigma$ upper limit is $4\times10^{-11}$~erg~s$^{-1}$cm$^{-2}$.
This is two orders of magnitude above the extrapolation of the LECS/MECS 
spectrum into the PDS bandpass.

A time history of the afterglow was constructed from all MECS data for the 2 to 
10 keV bandpass, see figure~\ref{figth}. The time bins were scaled 
logarithmically in order to optimize the signal-to-noise ratio in each bin. 
A power law function satisfactorily 
describes the data, the resulting power law index is 
$-1.5\pm0.2$ and $\chi^2_{\rm r}=0.8$ (3 d.o.f.). The trend in the afterglow
extends quite close to the WFC measurements of the burst (see 
figure~\ref{figth}). Therefore, in a next step we included the 2 to 10 keV flux 
as measured with WFC between 54 and 90~s after the trigger into the model.
The best-fit decay index then is $-1.35\pm0.03$ ($\chi^2_{\rm r}=0.8$ for 
4 d.o.f.). It is consistent with the index as determined from the afterglow alone.
We conclude that there is evidence that the afterglow
emission started during the tail of the burst event. Unfortunately, we cannot
substantiate this with spectral measurements because of insufficient
statistics.

There is WFC coverage from 24~h before till about 5~h after the burst, 
with intermissions. No emission was detected from the burst position outside
the 90~s interval after the trigger time. The $3\sigma$ upper limits are 
$1\times10^{-10}$~erg~s$^{-1}$cm$^{-2}$ for the pre-burst and 
$3\times10^{-10}$~erg~s$^{-1}$cm$^{-2}$ for the post-burst data (2 to 10 keV)

Greiner et al. (1998) reported a ROSAT observation that started 35~h after the 
end of the NFI observation and lasted 39~h (Greiner, priv. comm.). No X-ray 
afterglow was detected. For $N_{\rm H}=1\times10^{22}$~cm$^{-2}$, the ROSAT 
upper limit in 0.1-2.4 keV for the unabsorbed flux is 3 times the value 
predicted from an extrapolation of the MECS evolution. 

The decay function predicts a 2 to 10 keV fluence (i.e., integrated from 1~s 
after the trigger onwards) of $3\times10^{-6}$~erg~cm$^{-2}$. 
In the 2 to 700 keV band the afterglow fluence is 
one order of magnitude smaller than the fluence of the burst event.
The decay index is consistent with that of the R-band counterpart 
(Palazzi et al. 1998).
Also, it is very close to the average of -1.32 over the X-ray afterglows of 
GRB~970228 (Costa et al. 1997), GRB~970402 (Nicastro et al. 1998), GRB~970508
(Piro et al. 1998), GRB~970828 (Yoshida et al. 1998) and GRB~971214 (Heise
et al. 1998).

No flares or dips occur on time scales down to 10$^3$~s. The $3\sigma$ upper 
limit on upward fluctuations
on a 4000~s time scale is 70\% at the start and 240\% at the end of the
NFI observation. The data is marginally sensitive 
to detect flux increases like in GRB~970508 (about 200\%, 
Piro et al. 1998) or GRB~970828 (about 100\%, Yoshida et al. 1998). 

\section{Discussion}
\label{sectiondis}

The spectrum of the burst is exceptionally hard above the break energy during 
the whole event. Preece et al. (1998) studied the spectra of 126 bright bursts 
detected with BATSE during 1991 to 1997 above 28 keV,
with fluences comparable to or higher than that of GRB~980329. None of these
bursts shows a time-averaged $\beta>-1.51$. Another study of 22
GRBs detected with GRB sensors on board {\em Ginga} in the 2 to 400 keV
range shows a similar upper boundary of -1.50 (Strohmayer et al. 1998).
Our results indicate that in rare cases the radiative 
conditions applicable to high peak-intensity bursts can favor hard power-law 
components of the spectrum.

Of all 14 GRBs detected with WFC and GRBM, GRB~980329 is the brightest
in both X-rays and gamma-rays. The burst most similar to GRB~980329 is
GRB~970111 (Feroci et al. 1998). It has a similar ratio of GRBM to WFC 
photon count rate during the gamma-ray peak and its gamma-ray peak intensity and 
fluence are only $\sim$20\% smaller. Furthermore, the durations of both bursts are
similar and the gamma-ray time profiles are, by eye, of the same complexity.
The difference between both bursts is that the time profile of GRB~970111 shows
much more spectral evolution and is softer at high energies as indicated by 
the average for $\beta$ of $-2.13\pm0.03$ (Frontera et al. 1998b).

The behavior of the X-ray afterglow of GRB~980329 is quite similar to that of
afterglows of other GRBs, both in intensity and decay index (except for
GRB~970228 which is brighter). Therefore, in general there does not appear to 
be strong correlation between burst events and their X-ray afterglow.

GRB afterglow observations generally comply well with the relativistic blast
wave model in which relativistic electrons produce synchrotron radiation (e.g., 
Wijers, Rees \& Mesz\'{a}ros 1997; Sari, Piran \& Narayan 1998). In this model the 
spectral evolution is given by $F(\nu,t)\propto t^\delta\nu^\epsilon$ for a range
of frequencies and times which contain no spectral breaks. In each of two
states $\delta$ and $\epsilon$ are functions of only $p$, the power law exponent
of the electron Lorentz factor distribution. For the synchrotron
cooling stage of the blast wave, $\epsilon_{\rm s}=2\delta/3$. For the adiabatic 
cooling stage, $\epsilon_{\rm a}=2\delta/3-1/3$. For $\delta=-1.35\pm0.03$, 
$\epsilon_{\rm s}=-0.90\pm0.02$ and $\epsilon_{\rm a}=-1.23\pm0.02$. The measured
value of $\epsilon=-1.4\pm0.4$ complies well to both predictions with a weak
preference for adiabatic cooling for which $p=(-4\delta+2)/3=2.46\pm0.04$.

The brightness of the burst suggests the possibility of a relatively small 
distance. However, beaming effects may be important and may be related to the 
lack of strong spectral evolution of the burst and the relative hardness at high 
energies. A future redhift measurement of the 
proposed host galaxy (Djorgovski et al. 1998) and a continued monitoring of the 
radio scintillation (Taylor et al. 1998) may reveal more insight.


\acknowledgements
We are grateful to the staff of the BeppoSAX Scientific Operation Center,
the Mission Planning Team and the Science Data Center for their support in
obtaining and processing the data. The BeppoSAX satellite is a joint Italian 
and Dutch program.


\figcaption[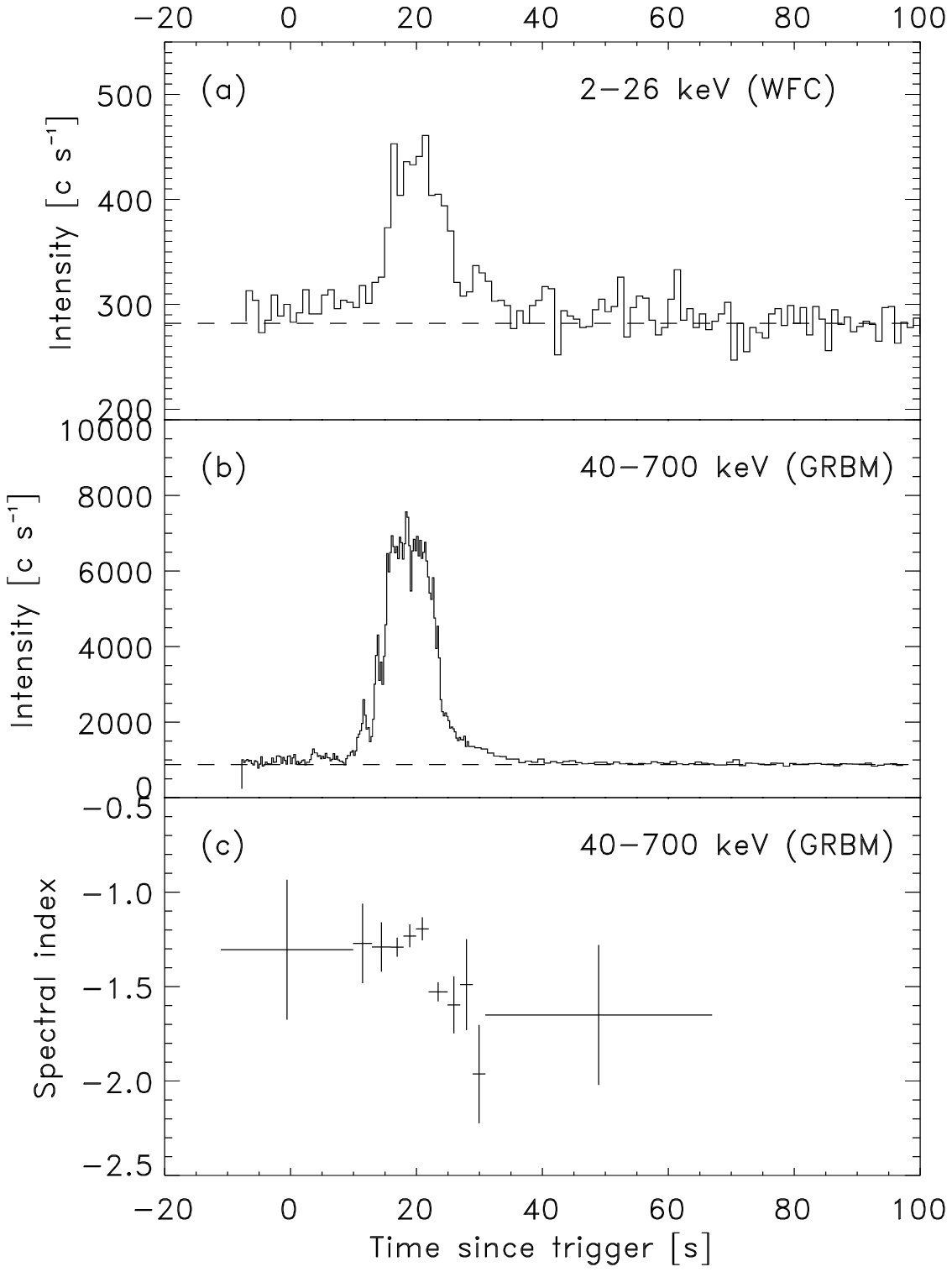]{
Time history of the burst itself as seen with WFC and GRBM, at a time 
resolution of 1~s (panel a) and 0.25~s resolution (panel b). Panel c 
refers to the the photon index of a power law function description of 
the GRBM two-channel data. The horizontal dashed lines refer to the 
background levels in the raw count rates. These are the averages as 
determined from the data beyond +90 s.
\label{figlc}}

\figcaption[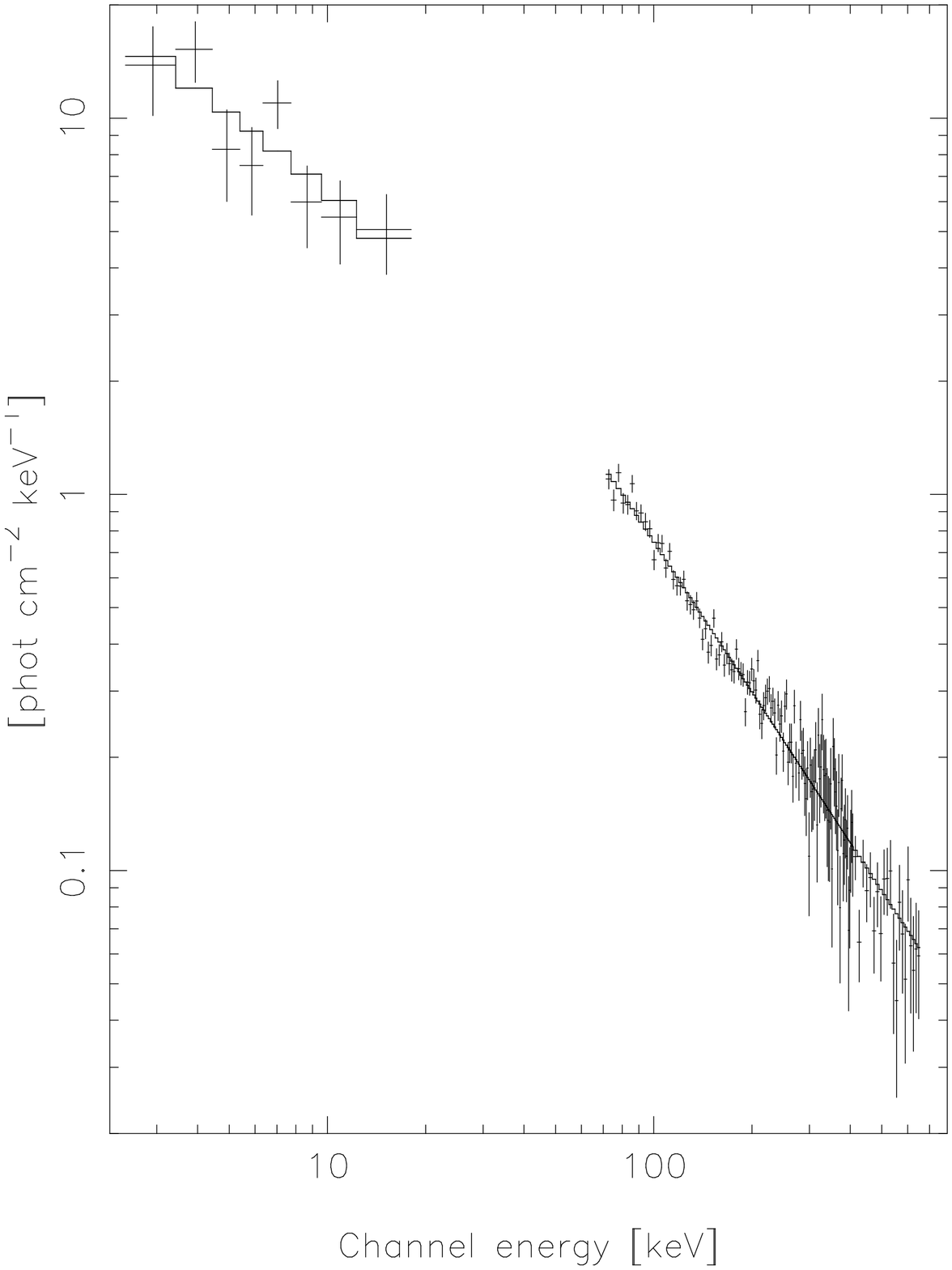]{
The photon spectrum of GRB~980329 as
measured with GRBM between -74 s and +54 s from the trigger time and
with WFC between +0 and +54 s. The drawn line refers to the best fit
"Band" spectrum (see text).
\label{figgrbspectrum}}

\figcaption[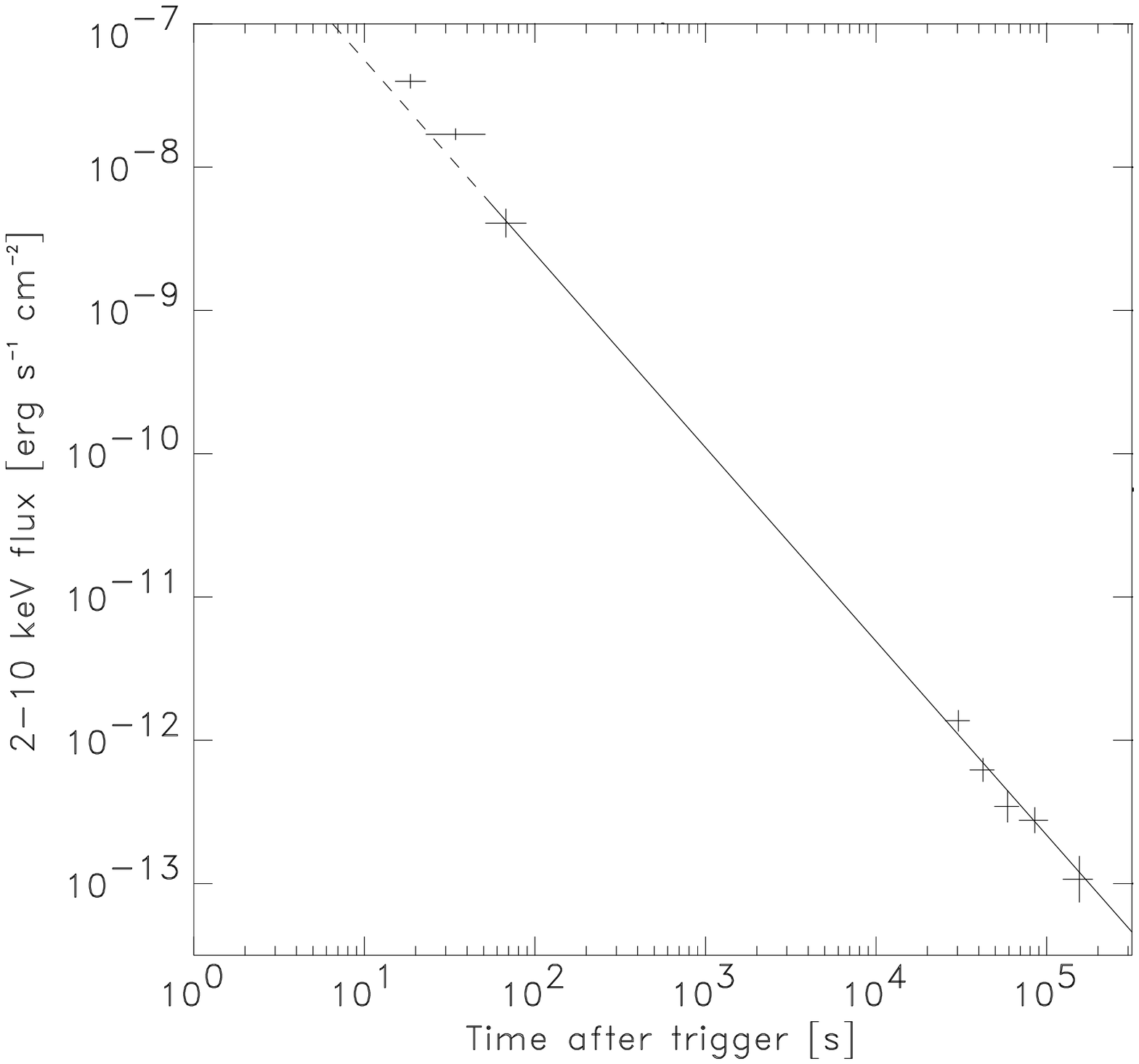]{
Time history of the X-ray afterglow, based on WFC (first 3 data points) and 
MECS data (last 5 points). The horizontal bars refer to the times of observation, 
vertical bars to the 1$\sigma$ errors. The first point is from 15 to 23 s
(see figure~\ref{figlc}) which is the brightest part of the burst.
The calibration of MECS c~s$^{-1}$ to 2 to 10~keV erg~s$^{-1}$cm$^{-2}$
is based on the overall LECS/MECS spectrum. The drawn line refers to a 
power law fit to the data (see text), the dashed line to an extrapolation.
\label{figth}}

\newpage
\setcounter{figure}{0}
\begin{figure}[t]
  \begin{center}
    \leavevmode
\epsfxsize=8.8cm
\epsfbox{grb980329_fig1.ps}
  \caption{
}
  \end{center}
\end{figure}

\begin{figure}[t]
  \begin{center}
    \leavevmode
\epsfxsize=8.8cm
\epsfbox{grb980329_fig2.ps}
  \caption{
}
  \end{center}
\end{figure}

\begin{figure}[t]
  \begin{center}
    \leavevmode
\epsfxsize=8.8cm
\epsfbox{grb980329_fig3.ps}
  \caption{
}
  \end{center}
\end{figure}

\end{document}